# Martini 3 Coarse-Grained Force Field for Carbohydrates

Fabian Grünewald[1*], Mats H. Punt[1*], Elizabeth E. Jefferys[2], Petteri A. Vainikka[1], Valtteri Virtanen[3], Melanie König[1], Weria Pezeshkian[1,5], Maarit Karonen[3], Mark S. P. Sansom[2], Paulo C. T Souza[4†], Siewert J. Marrink[1†]

[1] Groningen Biomolecular Sciences and Biotechnology Institute and Zernike Institute for Advanced Materials, University of Groningen, Groningen, The Netherlands

[2] Department of Biochemistry, University of Oxford, South Parks Road, Oxford, United Kingdom, OX1 3QU

[3] Natural Chemistry Research Group, Department of Chemistry, University of Turku, FI-20014 Turku, Finland

[4] Molecular Microbiology and Structural Biochemistry, UMR 5086 CNRS and University of Lyon, Lyon, France

[5] The Niels Bohr International Academy, Niels Bohr Institute, University of Copenhagen, Copenhagen, Denmark

* shared first authors

† corresponding author

*Supporting Information Placeholder*

**ABSTRACT:** The Martini 3 force field is a full re-parametrization of the Martini coarse-grained model for biomolecular simulations. Due to the improved interaction balance it allows for more accurate description of condensed phase systems. In the present work we develop a consistent strategy to parametrize carbohydrate molecules accurately within the framework of Martini 3. In particular, we develop a canonical mapping scheme that decomposes arbitrarily large carbohydrates into a limited number of fragments. Bead types for these fragments have been assigned by matching physicochemical properties of mono- and disaccharides. In addition, guidelines for assigning bonds, angles, and dihedrals are developed. These guidelines enable a more accurate description of carbohydrate conformations than in the Martini 2 force field. We show that models obtained with this approach are able to accurately reproduce osmotic pressures of carbohydrate water solutions. Furthermore, we provide evidence that the model differentiates correctly the solubility of the poly-glucoses dextran (water soluble) and cellulose (water insoluble, but soluble in ionic-liquids). Finally, we demonstrate that the new building blocks can be applied to glycolipids, being able to reproduce membrane properties and to induce binding of peripheral membrane proteins. These test cases demonstrate the validity and transferability of our approach.

## 1 – Introduction

Carbohydrates (sugars) are an important class of biomolecules. They play an active role in cell biology as they are, for example, part of the cell metabolism[1], or signaling pathways[2]. In addition, they are structural building blocks for many bio-macromolecules such as polysaccharides, glycosylated proteins and lipids, and nucleotides. Furthermore, in research for sustainable materials carbohydrates are a key factor as they can be obtained from renewable stock.[3] Therefore, simulating these molecules in complex systems by molecular dynamics (MD) is of high interest to a wide audience of researchers. MD studies can give near atomistic resolution of processes not possible to capture with experimental techniques and therefore often complement experimental studies.

Due to the limits in special temporal resolution of models representing all atoms explicitly, so-called coarse-grained (CG) models are often used in MD simulations. In CG models, several atoms are grouped into one effective interaction side. This greatly increases the simulation speed and reduces computational costs. Among the most popular CG models for (bio-) molecular dynamics is the Martini model.[4,5] The Martini model has been widely applied across many fields ranging from biomolecular science to material science.[6–9] In the Martini model,[5] about four heavy atoms are grouped into one interaction center, called bead. The interactions between beads represent the nature of the underlying chemical groups; the strength of the interaction is selected from a discrete set of LJ interactions by reproducing thermodynamic data mostly the free energies of transfer between water and different organic solvents. In addition to the regular Martini beads, smaller bead sizes (S- and T-beads) are used for groups that are represented at higher resolution such as aliphatic or aromatic ring fragments.[5,10]

Within the framework of the previous version of the Martini model (i.e. version 2), several parameters for carbohydrates have been developed and successfully applied.[11–26] However, Martini 2 has several pitfalls when it comes to parameterization of molecules, which lead to unphysical behavior.[27] This was especially apparent for the carbohydrate model. As pointed out by several authors carbohydrates in Martini 2 tend to largely overestimate the self-aggregation propensity.[12,13] Although some of these problems could be alleviated by either increasing the interaction strength of carbohydrates with water[13], or by replacing regular bead types with small beads[18], these solutions were *ad-hoc* and did not resolve the underlying imbalance of the bead interactions. To overcome these deficiencies the third edition of the Martini force field comprises a complete reparameterization of the original force field. Rebalancing of the non-bonded interaction as well as extended verification against physicochemical reference data, leads to an improved description of previously problematic molecular interactions.[5,10,28–31]

In the present work we develop a consistent strategy to parametrize arbitrary carbohydrate molecules accurately within the framework of Martini 3. In particular, we develop a canonical mapping scheme that decomposes large carbohydrates into mono- and disaccharides, which are parameterized based on matching physicochemical reference properties and atomistic reference simulations. To facilitate application of this scheme, automatic mapping from all-atom simulations is implemented in the fast-forward program[32] for all carbohydrate fragments considered. In addition, we propose guidelines for assigning bonds, angles, and dihedrals to allow for a more accurate description of carbohydrate conformations than in the Martini 2 force field. At the moment bonded interactions for specific complex carbohydrates need to be mapped from an atomistic reference simulation unless the specific compounds are presented in this paper. Generic bonded parameters are subject to a forthcoming publication.

The remainder of this paper is organized as follows: First we present the parametrization strategy of carbohydrates starting with monosaccharides (Section 2.1) and subsequently extending to disaccharides and more complex carbohydrates (Section 2.2). Afterwards, we validate the transferability of our approach by demonstrating that we can accurately model four example systems, previously impossible to consistently model with Martini 2. In particular, reproduction of osmotic pressures of monosaccharides (Section 3.1), solution and solubility behavior of two polysaccharides namely dextran and cellulose (Section 3.2 & 3.3), and glycolipid mediated binding of peripheral membrane proteins (Section 3.4). Finally, we discuss limitations of our new carbohydrate modeling strategy and conclude.

## 2 – Parametrization of carbohydrates with Martini 3

We derived parameters for Martini 3 carbohydrates following the general rules for creating Martini models as outlined in the main parameterization paper.[5] However, we aimed at not only deriving the optimal parameters for the specific carbohydrates considered, but also casting them in a consistent framework as much as possible such that we obtain a generalized strategy for modeling arbitrary carbohydrates.

### 2.1 – Monosaccharides

***Scope.*** Carbohydrates as a molecule class display a large heterogeneity in size, structural connectivity and isomerization states. Most biologically and technologically important carbohydrates are either hemiacetal monosaccharides or formed by condensation reactions of hemiacetal sugars. The hemiacetal monosaccharides exist mostly as six membered carbon rings (pyranoses) or five membered carbon rings (furanoses). These monosaccharides can undergo enantiomerization reactions switching between an open form and a ring-closed from in solution. However, the total fraction of ring-open structures is typically very low. Therefore, in this paper we only consider ring-closed monosaccharides. Furthermore, the C1 carbon, which is called anomeric carbon, is chiral. This chirality of hemiacetal monosaccharides also causes a specific type of isomerization called anomerization. Depending on the position of the alcohol group connected to the anomeric carbon a carbohydrate either has a α- (axial alcohol) or β- (equatorial alcohol) conformation. Within the Martini 3 carbohydrate model we do not distinguish between the anomers in the case of monosaccharides. As shown in the Table S1 the geometry (i.e. bond lengths) are so similar that we can treat these molecules as one class as was done in the previous Martini carbohydrate models.

***Mapping.*** The mapping describes which atoms at the all-atom level are represented by a single bead in the CG model. The mapping choice determines all subsequent model choices and therefore requires careful consideration. In general, in the context of Martini 3 one maps 2-5 heavy atoms into one bead, where the number of mapped atoms and their connectivity determines the bead-sizes. All carbohydrate mappings are derived obeying the following three base rules, which are aimed at making the mappings transferable and consistent across also complex carbohydrates (Figure 1a).

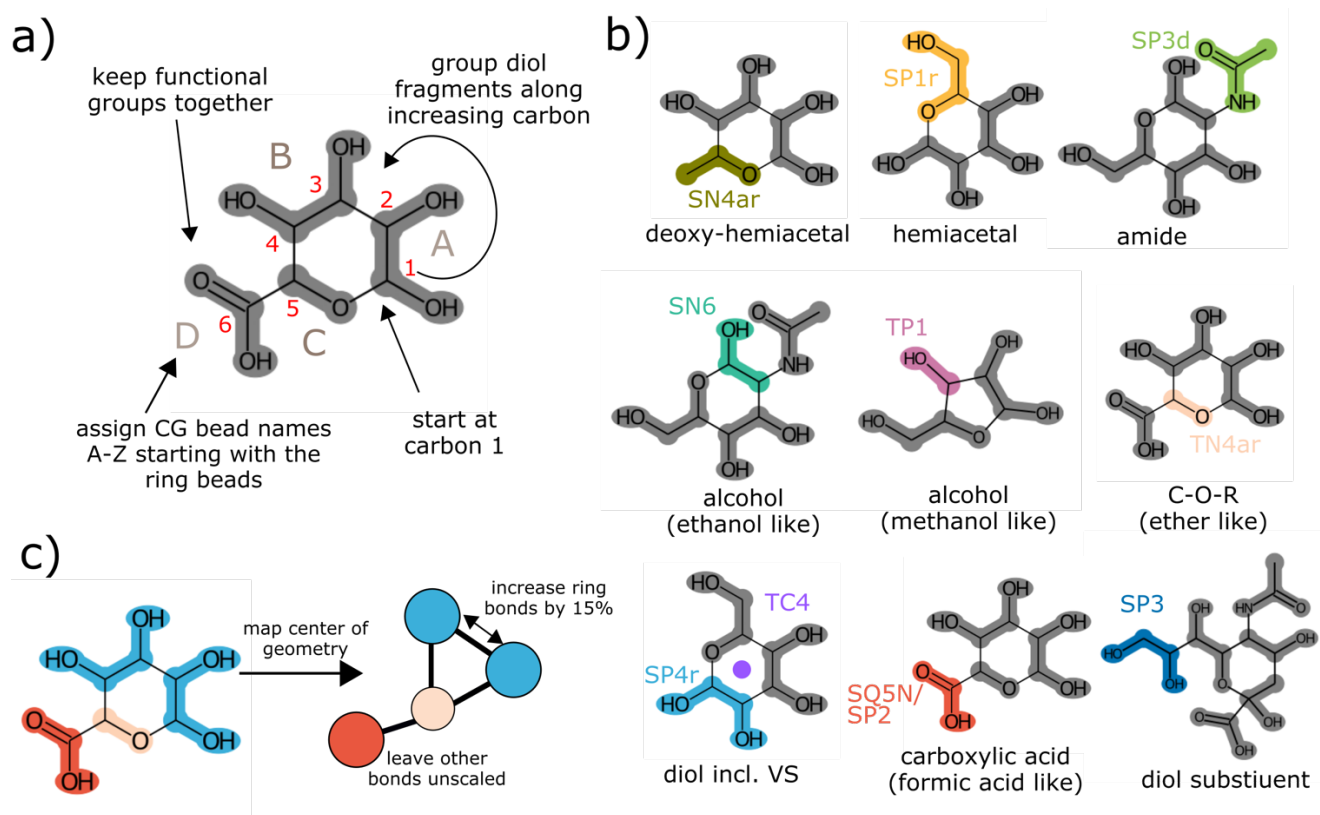

**Figure 1. Parametrization strategy for monosaccharides. a) Systematic mapping scheme; b) Bead assignment for all fragments found in monosaccharides; c) Design principle for bonded interactions.**

A mapping: 1) maximizes the number of diols assigned to a single bead therefore maximizing the number of 4:1 mapped atoms; 2) keeps functional groups together as much as possible; 3) starts at the anomeric carbon and proceeds counterclockwise for grouping fragments. The first rule recognizes that the most commonly found fragments in sugars are diols and the hemiacetal group. Having as many similar fragments as possible simplifies the model and allows to find a good bead type for that fragment across the many test cases. The second rule is needed in cases where substituent groups are presents and supersedes rule 1, if needed. For example, in D-glucuronic acid (Figure 1a), the acid substituent group is kept together making the ring-fragment smaller and in the case of Neu5Ac the three side chains are also kept together (Figure S1). Thus, the ring fragment becomes a 3-1 mapping. The third rule ensures that equivalent fragments are generated for the different sugars and makes a canonical naming scheme possible. To simplify handling and analysis of our model we have developed a canonical naming scheme. The ring-beads in carbohydrates are named A, B, C, where the A bead is the first bead and always includes the anomeric carbon according to the previously defined mapping direction and the C bead always includes the ether oxygen. Substituents are named after letters in the alphabet in order in which they are attached to the main ring beads. For example, in the case of glucuronic acid we have one substituent named D.

*Bonded interactions.* As the monosaccharides are rigid triangles at the CG level, we decided to model them using constraints. The constraint length was derived by mapping the center of geometry (COG) of the atomistic reference structure to the CG representation following the previously derived mapping scheme. One of the key problems of sugars in the old Martini 2 model was the overestimated aggregation propensity.[12,13,27] As analyzed by Alessandri and coworkers, this is partially caused by too short bond lengths and a bad representation of the molecular volume at the CG level.[27] Thus in Martini 3 it is recommended to match the molecular volume of the atomistic structure as closely as possible.[27] To assess the molecular volume, we conducted atomistic simulations of 11 monosaccharides, and 3 disaccharides at the all-atom level using the most popular force fields for sugars (i.e., Glycam06h[33], Charmm36[34], and GROMOS[35]). Based on these simulations the SASA was computed as outlined in the method section and compared to CG SASA values. As the SASA depends only on the bead size we assigned bead-types of an appropriate size, based on the mapping guidelines of the Martini 3 model. As most of the fragments are 4:1 mapped that display branched moieties or 3:1 mapped linear groups, we considered mostly S-beads in accordance with the Martini 3 mapping guidelines. For some of the remaining 2:1 mapped fragments, the appropriate class of T-beads was used instead. Figure 2a shows the comparison of the Martini 3 SASA values against atomistic simulation data. The models, which are obtained by simply mapping the center of geometry (red symbols), underestimate the molecular volume significantly. Overall deviations for the unscaled model are of the order of 8%. Thus, it stands to reason that this approach leads to similar problems as observed in Martini 2.[27] To further elucidate the problem, we computed the Connolly surfaces[36] of the molecules involved. An example is shown for glucose in Figure 2 b-e. One can clearly see that the unscaled coarse-grained surface (blue) does not match the atomistic reference surface (red). To improve the agreement with the molecular volume we followed the approach suggested by Alessandri *et al.*[27] and increased the bond lengths of the beads forming the sugar rings. The scaling approach has also been used previously to improve interactions of PIP lipids, which contain a carbohydrate head group.[30] To keep our model transferable to carbohydrates not considered, we explored a compound independent scaling factor. A uniform scaling of 15% over the COG mapped distances was found to greatly improve the agreement of the SASA (orange data-points Fig. 2a) and at the same time be applicable to all monosaccharides. Also, the Connolly surfaces show a better agreement (Figure 2 d,e). Now the red and the blue surface align well for most parts of the molecule.

*Bead choices.* Non-bonded interactions are assigned from a discrete set of interaction levels referred to as bead-type by selecting those types that reproduce best the available physicochemical reference data. In this particular study we selected the bead types by matching the free energies of transfer from octanol to water for 11 monosaccharides. We note that only experimental values for glucose are available[37–39] (-17.52 ± 1). Thus, we set out to measure the remaining partition coefficients experimentally ourselves. The value obtained for glucose (-17.81 ± 0.5) matches the previously published values well, giving confidence in the choice of experimental method. Table S2 summarizes the experimentally determined partition coefficients.

The bead assignments were then optimized under the constraint that the same fragments must have the same bead type to be consistent with the building block approach of Martini. For example, inositol consists of three diol units. Thus, all beads in inositol must have the same type and it fixes the choice for the diol bead fragment.

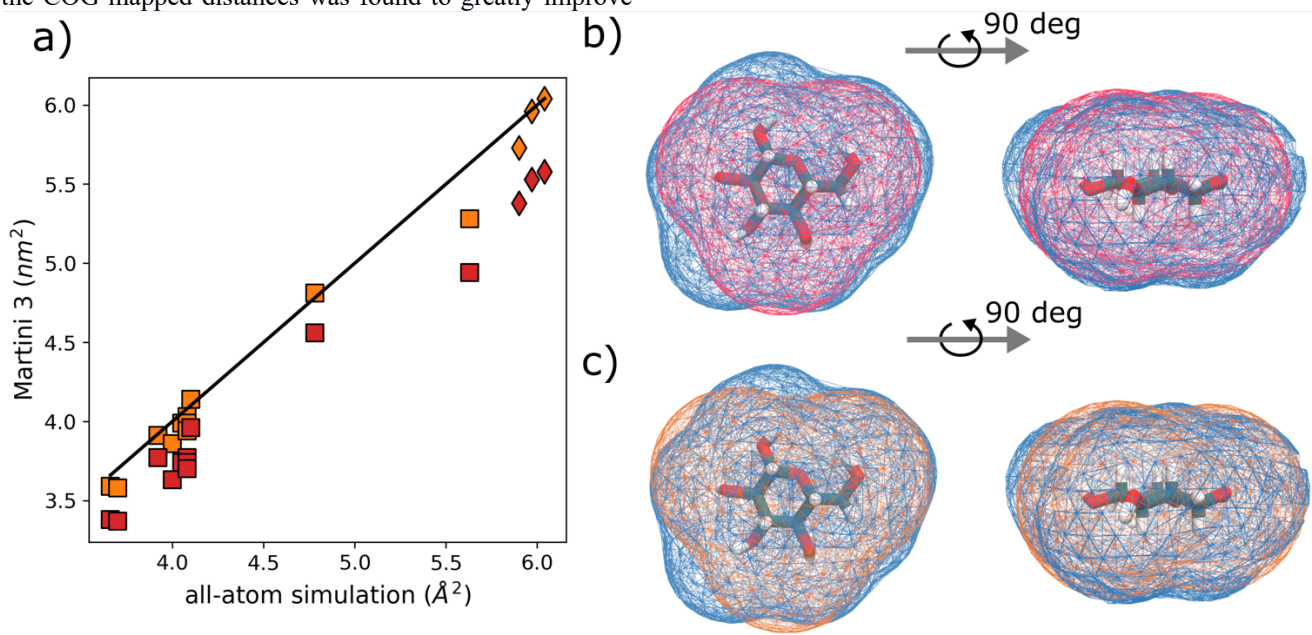

**Figure 2. Molecular shape optimization. a)** Solvent accessible surface area (SASA) compared between atomistic reference simulations and Martini 3 with unscaled bonds (red) and scaled bonds (orange), for monosaccharides (squares) and disaccharides (diamonds); **b)** Connolly surfaces for glucose comparing atomistic (blue) to Martini 3 (red) before scaling the bonds **c)** Connolly surface for glucose after scaling the bonds comparing atomistic (blue) to Martini3 (orange)

Starting with that assignment the choices for the other fragments could be optimized. Figure 1b shows the final bead assignments for the monosaccharides considered. Figure 3 shows the correlation of the experimental versus coarse-grained free energies of transfer. We note that the mean-absolute error across all monosaccharides is only 1.5 kJ/mol which we consider excellent. For comparison, it is about the same as the average error in transfer free energy for the small molecules considered in the Martini 3 parametrization, which is 2.0 kJ/mol.[5]

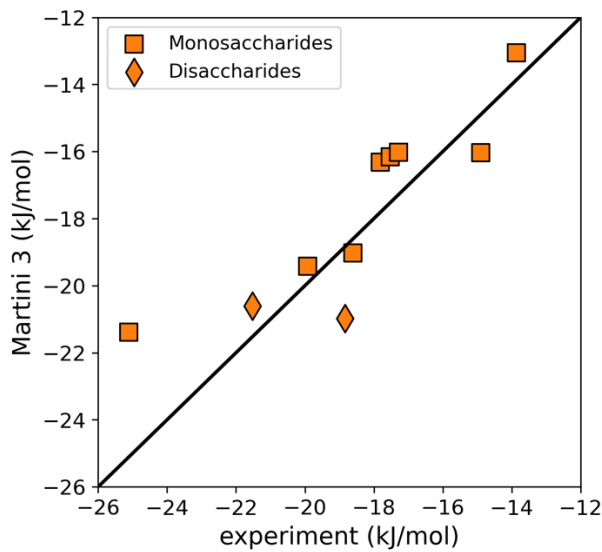

**Figure 3. Free energies of transfer.** Octanol - water free energies of transfer were computed using the Martini 3 model developed here and compared to either newly experimentally measured or existing literature values for monosaccharides (squares) and disaccharides (diamonds). See Table S2 for actual data. The error for all points was less than 0.4 kJ/mol.

In addition to assigning a bead type for each fragment in the three membered ring our model also contains a virtual interaction site (VS), which is placed at the center of geometry of the ring. The VS is a mass-less interaction site that has the bead type TC4 across all monosaccharides. Note that all bead assignments were done with the TC4 interaction site present. This VS helps to reproduce interactions with aromatic groups — through so-called ring stacking[40–42] – which in Martini needs to be captured through a hydrophobicity component to the interaction. In addition, the extra TC4 bead avoids excessive mismatch of number of non-hydrogen by using S-beads representing 4 branched non-hydrogen atoms. As described in the Martini 3 guidelines, the maximum mismatch should be 1 non hydrogen atom for each 10 non hydrogen atoms mapped by CG beads.[5] Such an approach has already been used successfully in parametrization of phosphatidylinositide lipids with Martini 3.[30] In the Supporting Information we assess the effectiveness of the virtual site by computing the potential of mean force (PMF) profiles between indol and glucose (Figure S2).

2.2 Disaccharides and more complex carbohydrates

*Mapping.* The mapping of disaccharides directly follows from the mapping of the monosaccharides, that is each constituting monosaccharide is mapped following the rules outlined above. The only thing to consider is the problem of where to separate the disaccharide into its constituting monomers. We adopt the same approach as previously used by the CHARMM-GUI glycan reader[43], considering the connecting atoms to belong to the monosaccharide with the higher bead number at the CG level. For example, in the case of lactose ($C_{12}H_{22}O_{11}$), which is an alpha-1,4 linkage of glucose and galactose (see Figure 4a), the connection is between the B and A bead of the CG model and between carbon 5 and 1 of the atomistic model. We consider the ether fragment to belong to the B bead fragment. This mapping allows to obtain transferable mappings between all disaccharides and since it is consistent with the CharmmGUI convention it allows automatic forward and backward mapping using already existing tools such as fast-forward[32] or backwards[44]. The complex carbohydrates and polysaccharides are mapped following the same principle, which also holds for branched carbohydrates such as the GM1 lipid. This scheme also allows Martini 3 carbohydrates to keep a building block approach, which was a concern in the previous Martini 2 model, where the CG geometry of the sugar rings needed to change from monosaccharides (triangular topology) to oligosaccharides models (linear).[45]

*Bonded interactions.* Bonded interactions are derived following the same strategy as for the monosaccharides that is mapping and matching the underlying atomistic reference distributions. To best represent the conformational space underlying carbohydrate oligomers and polymers we define all angles spanning the bond between two monosaccharide repeat units as well as one dihedral controlling the rotation around the glycosidic bond (Figure 4b). In case more than two monosaccharide repeat units are connected, an additional dihedral angle is introduced that spans three repeat units (Figure 4c). This dihedral angle defines the relative orientation of the n and n+2 residue with respect to the plane spanned by the n+1 residue.

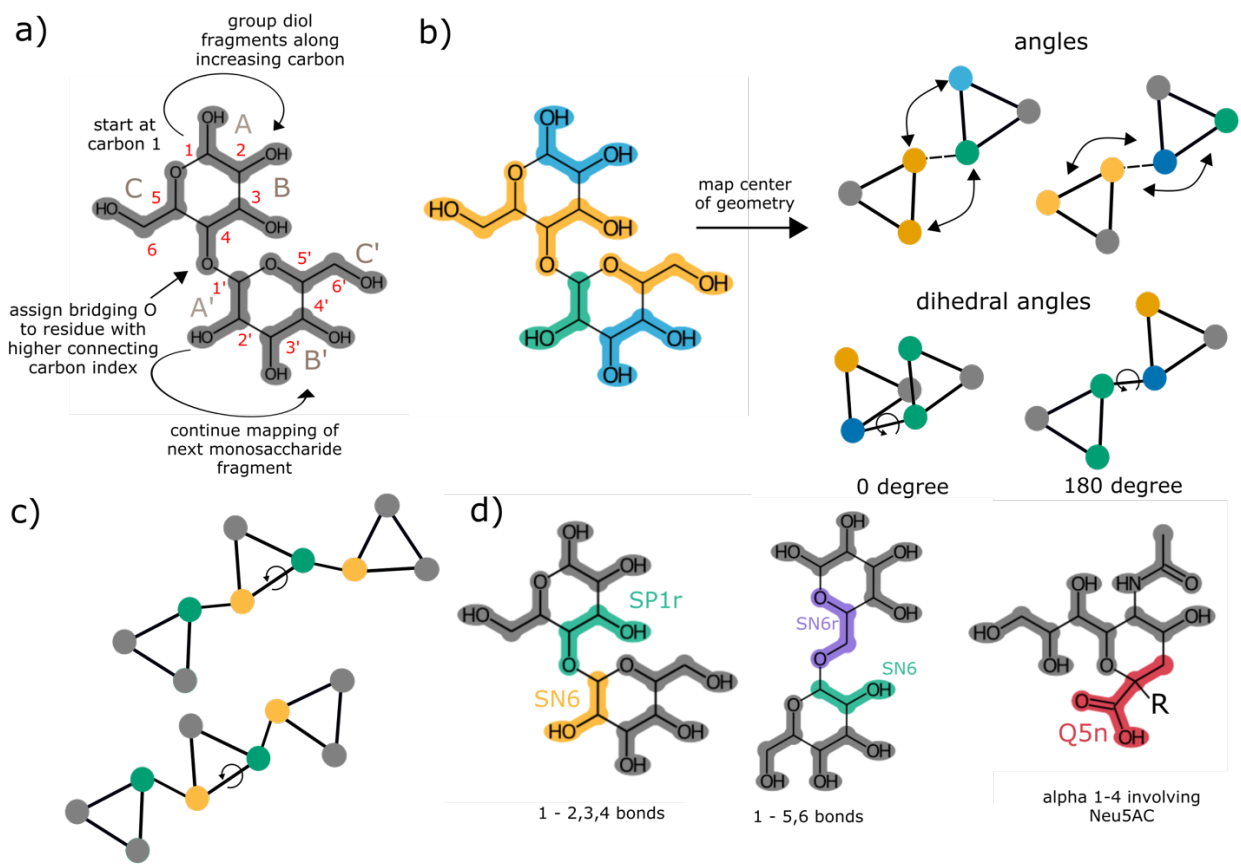

**Figure 4. Parametrization strategy for oligo- and polysaccharides. a)** Systematic mapping strategy for complex carbohydrates; **b)** Angles and dihedrals introduced between two linked monosaccharide fragments; **c)** Dihedral angle introduced between three consecutive monosaccharide fragments; **d)** Bead assignment for all fragments not found in monosaccharides.

We notice that this dihedral is important especially for longer carbohydrates as it relates to the stiffness of the polysaccharides. Polysaccharides formed by condensation can either have a α or β based bonds. We noticed that the difference in bond length at the CG level is significant between an α or β bond due to the relative positioning of the two rings with respect to each other. Therefore, in our model we distinguish explicitly between the two anomers when found in a poly- or oligo-saccharide. In the case of dextran polymer to allow for better matching of the underlying AA distributions and improved stability, we use 3-bonded neighbor exclusions. However, for all other models only the 1-bonded neighbors are excluded as is standard in Martini lipids. Furthermore, for angles that are covered by any dihedral potential, the restricted bending potential introduced by Bulacu *et al.*[46] is used to reduce instabilities from angles becoming co-linear. As for the monosaccharides, we have assessed how well our model represents the molecular volume by computing SASA values for three disaccharides lactose, sucrose, and trehalose (Figure 2a). The correlation with the atomistic SASA values is similarly good as for the monosaccharide case given the bond scaling is retained. However, it should be noted the connecting bonds between two monosaccharide repeat units are left unscaled. This minimal number of bonded interactions provides a good representation of the all-atom conformations and also leads to a numerically stable model. Moreover, it improves in selectivity and functionality compared to the Martini 2 carbohydrates, which treated glycosidic bonds indiscriminately and even had problems modelling flexible α-1,6 linkages[11]. Note that bonded parameters for other complex carbohydrates (especially dihedrals) should be mapped from an atomistic reference simulation. Using generic bonded interaction to combine arbitrary carbohydrates is beyond the scope of the current paper and will be discussed in a forthcoming publication.

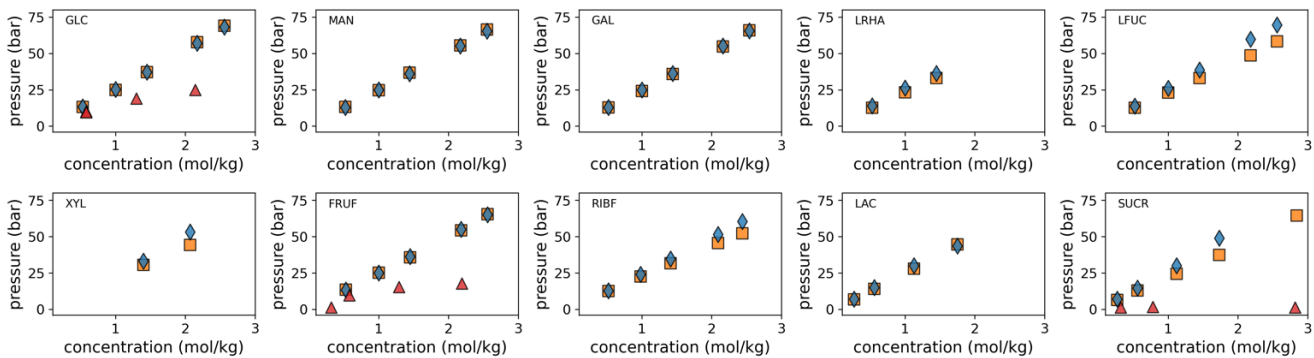

**Figure 5. Osmotic pressure of carbohydrate solutions.** Osmotic pressure for ten different carbohydrates measured from simulations using the presented Martini 3 model (orange) and the regular Martini 2 model (red) in comparison to experimental data (blue diamonds) collected from various sources[47–50]. Mono- and disaccharide codes are indicated with each panel: GLC=D-glucose, MAN=D-mannose, GAL=D-galactose, LRHA=L-rhamnose, LFUC=L-fucose, XYL=D-xylose, FRUF=D-fructose, RIBF=D-ribose, LAC=lactose, SUCR=sucrose. The error on CG osmotic pressures was less than 1 bar and thus smaller than the symbols shown.

***Bead choices.*** Bead types of the fragments which are equivalent in both mono- and disaccharides are retained following the building block spirit of Martini. Therefore, the only beads types, which need to be defined are those involved in the glycosidic bond. As shown in Figure 4d, hexose bonds can be collected into two groups based on the newly generated fragments. One group contains the 1-1, 1-2, 1-3, and 1-4 bonds. The appropriate bead type of this fragment, SP1r, is directly taken from the monosaccharides (cf. Fig 1b). However, we further validated this choice by reproducing the free energies of transfer between octanol and water of the disaccharide's trehalose and sucrose. Deviations for both compounds were acceptable with errors of 0.5 kJ/mol for trehalose and about 3.0 kJ/mol for sucrose. The other group contains 1-5 and 1-6 glycosidic bonds, in which case another bead type needs to be assigned.

The new bead is similar to the hemiacetal fragment but for the change of one OH group to an ether group. Such a change will likely result into less strong self-interactions, which is captured by the SN6r bead type having one level less strong self-interaction. Finally, in case of N-acetylated neuraminic acid attached via a 1-4 bond (Figure 4d last panel), we group the carboxylic acid together with the remaining carbon fragment in order to avoid a 2:1 fragment being generated with a short bond length. As a consequence, a new large bead is used instead of the two smaller beads. The bead type was determined to be the standard carboxylic acid bead from the original Martini 3 publication. We note that as a result of the generalized mapping scheme as well as the fact that biologically relevant carbohydrates are confined to certain linkages, with this small number of new bead-types almost all biologically relevant sugars can be constructed.

### 3.0 – Validation

In order to demonstrate the transferability of the model and assess if the model improves on the multiple issues of the Martini 2 model, we analyzed a number of test cases, considering four different target systems.

### 3.1 Osmotic pressures

The major drawback of the Martini 2 carbohydrate model is the overestimation of the aggregation propensity.[12,13,27] To quantitatively asses the aggregation propensity of solute molecules in solution the osmotic pressure is frequently computed as function of the concentration. An osmotic pressure lower than experiment is indicative of too strong aggregation propensity. This procedure has been applied to reparametrize both CG and AA force fields for carbohydrates and other molecules.[13,34,51,52] To compute the osmotic pressure for our carbohydrates we have adopted the procedure originally proposed by Luo and Roux.[53] The molal concentrations were determined from the box density after preparing it at a certain molar concentration. Since experimental measurements are generally reported in molal units, we considered this approach to be more accurate at higher concentrations. Figure 5 shows the osmotic pressures for eight monosaccharides and two disaccharides in the concentration range from 0-2.5 molal. The scaled bond model (orange squares) shows an excellent agreement with the experimental data (blue diamonds) in the lower concentration range (<1.5 molal) across all carbohydrates. This already presents a significant improvement over the Martini 2 model for which data was available for only three carbohydrates (red triangles).

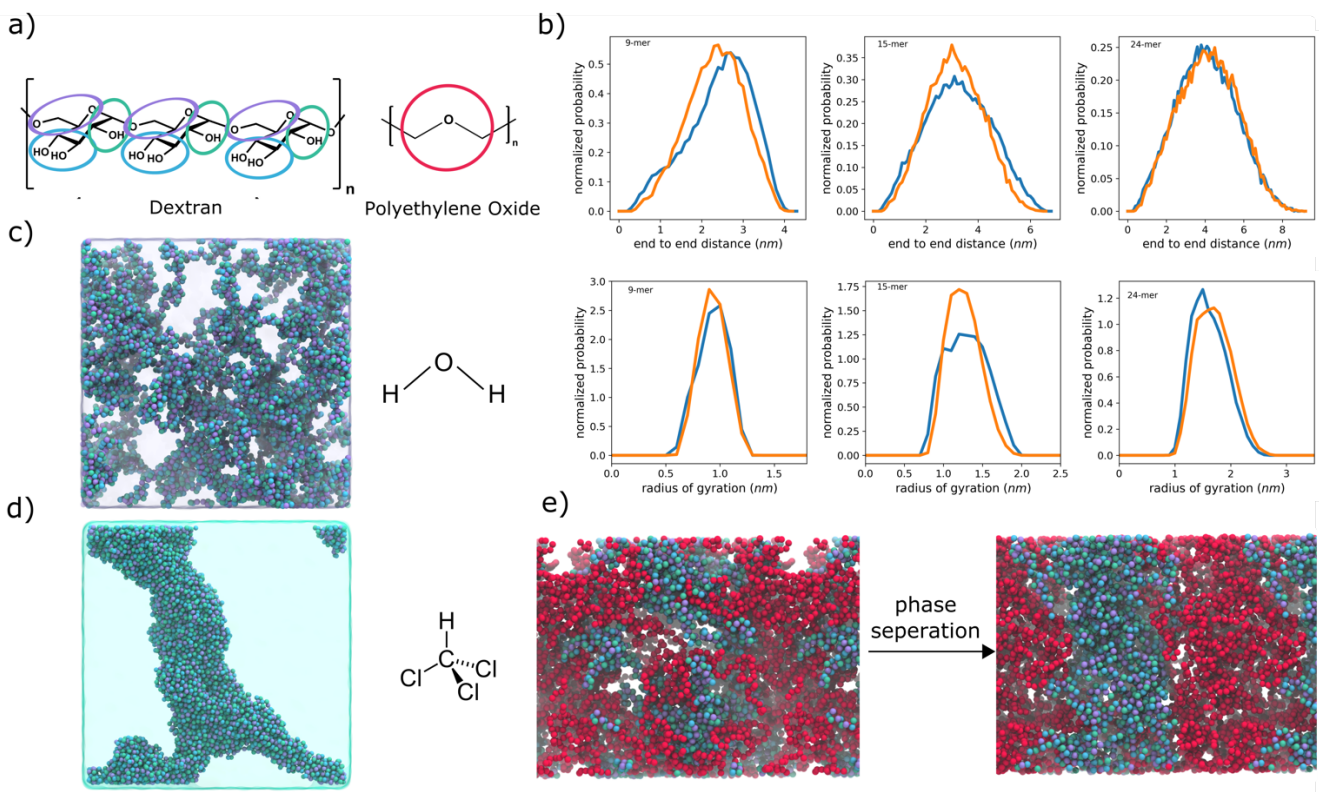

**Figure 6. Solution properties of dextran. a)** Mapping of dextran and polyethylene oxide (PEO) at the Martini level. The colors of dextran correspond to the bead-types as found in Figure 1 and 4. **b)** Radius of gyration and end-to-end distance of dextran oligomers with 9, 15 and 24 repeat units from all-atom CHARMM36m[43,54] simulations (blue) and the new Martini 3 model (orange). **c)** Snapshot of aqueous solution of dextran (50 repeat units). **d)** Dextran globule formed in chloroform, a non-solvent, starting from dispersed polymers. **e)** Aqueous solution of dextran (65 repeat units) and PEO (180 repeat units) at the beginning of the simulation in the fully mixed initial state (left) and after 3 microseconds of simulation (right). Solvent is omitted for clarity.

For the higher concentration range (>1.5 molal), we see that our model follows the overall trends well but shows some deviations, in particular, for ribose, sucrose, fucose, and xylose. The lower pressures observed for these carbohydrates suggests some remnants of the stickiness problem to be still present. However, overestimated solute-solute interactions in concentrated solutions are not unique to CG force fields like Martini. Even popular atomistic force fields such as CHARMM36 or GLYCAM06 have been shown to significantly underestimate osmotic pressures and therefore exaggerate aggregation in simulation of carbohydrates in water. Agreement with experimental data is especially bad at higher concentrations (>2.5 molal).[34,51,55] The reported deviations for these atomistic force fields are similar or even much worse than the deviations observed for our CG carbohydrate model.

Keeping this in mind, we conclude that our model reproduces the osmotic pressure very well overall and constitute a significant improvement over the Martini 2 carbohydrate models. In addition, we note that the accuracy of our model is comparable to default atomistic force fields. Thus, we consider the match to be good enough.

### 3.2 Solution properties of dextran

Dextran is a branched polysaccharide, consisting of $\alpha_{1,6}$ glucose units with $\alpha_{1,3}$ connected glucose units branching off it.[56,57] Unlike other poly-glucoses such as cellulose or amylose, dextran is fully water soluble even at high molecular weight fractions.[58] Here we investigate the solution properties of dextran in water to demonstrate that our model is not only capable of reproducing properties of mono- and disaccharides but also properties of complex polysaccharide solutions. Whereas high molecular weight dextran is usually highly branched, lower molecular weights (with degrees of polymerization below 100) typically display only in the order of 5% or less branching.[56,57] Since we mostly utilize such lower molecular weights, all dextran used here is linear and has no branches. Dextran bead-types (Figure 6a) were assigned based on the previously presented concepts and the bonded parameters were obtained by matching atomistic CHARMM36m[43,54] simulations. Subsequently we investigated the dilute solution conformations of three oligomers differing in the degree of polymerization (DoP 9, 15, 24). Figure 6b shows distributions of the radius of gyration as well as the end-to-end distance for the all-atom model (blue) and the Martini 3 model (orange). We note that the agreement is excellent between both sets of simulations for both metrics. The end-to-end distance is typically related to polymer stiffness via the persistence length. Seeing that our end-to-end distance distributions agree well with the AA model, this indicates that our persistence length at least for dextran oligomers is very close to what is predicted by the AA model. In contrast, the scaling of the radius of gyration is related to the dilute solution thermodynamics via for example Flory theory[59]. Being able to match the AA distributions closely indicates that we capture the solution thermodynamics well, suggesting that our model has a good balance between self and water interactions. We continued to investigate further, if dextran is still soluble at higher weight fractions. Figure 6c shows a snapshot of a dextran (DoP 50) solution at 10w/w%, after 5 microseconds of simulation. Clearly no phase separation or strong aggregation is visible. In the Supporting Information (Figure S3), we further show the radial distribution functions (RDFs) for the polymer-polymer and polymer water interactions. Both indicate that the dextran is fully solubilized in water. To set these results into perspective, we also simulated dextran in chloroform, which is a known non-solvent[58], for 5 microseconds. Already within a few hundred nanoseconds the polymers all aggregate into a periodic cluster (Figure 6d). Note that the concentration is the same in both simulations. The RDF for the polymer and polymer chloroform interactions (Figure S3) also shows the increased aggregation and depleted solvent interaction. We conclude our dextran model is fully water soluble and insoluble in chloroform, as expected for these concentrations and molecular weights.

Our final test-case involves an aqueous system of dextran and polyethylene oxide (PEO). Dextran is known to phase separate from PEO in a ternary mixture within water forming an aqueous two-phase system (ATPS) via liquid-liquid phase separation (LLPS).[60,61] ATPSs are important both in biomedical applications[62], for instance, microfluidic separation and in biological research[60], where they are used as compartmentalizing cytosol mimetic. Martini 3 has previously been shown to be able to capture LLPS of biomimetic compounds[63]. Thus, our model should also be capable of simulating those systems accurately. To this end, we generated a mixed PEO (DoP 180)-dextran (DoP 65) system using polyply[64]. We note that the used molecular weights and concentrations have previously been reported to from an ATPS.[60] Figure 6e shows the system at the start and end of a 2 μs simulation. Clearly the system has phase separated from the initial mixed state into a dextran rich phase and a phase enriched in PEO. Analysis of density profiles along the cartesian z-axis supports this conclusion (Figure S4). Overall, we showed that our dextran model matches solution conformations of atomistic simulations well, dissolves up to high weight fractions in water, does not dissolve in chloroform and forms an ATPS with PEO. All these observations are consistent with experimental data and demonstrate the validity of this carbohydrate polymer.

### 3.3 Solution properties of cellulose

In the previous two sections, we have shown that our Martini model produces carbohydrate molecules, which are water soluble and do not suffer from the same aggregation effects as seen for Martini2. However, not all carbohydrates are water soluble. For example, cellulose is famously known to be insoluble in water. In order to verify that our balance of interactions is reasonable and does not favor water soluble systems too much, we assessed solution properties of short cellulose analogs. In particular we simulated poly-(β-1,4) glucoses with a DoP of 50 (Figure 7a). At these lengths it is known to from stable crystals in water that do not solubilize.[65] We started by building a system in a perfectly mixed state and random chain conformations and simulated it for 3 us. Figure 7b shows the starting configurations as well as the last frame of the simulation. Clearly the cellulose analog starts aggregating even forming small fiber like structures. We further investigated if pre-build crystals of cellulose are stable. To this end, a cellulose crystal (1β) with was build using cellulose-builder[66], solvated and run for 2 microseconds. Reassuringly, the fibril remains stable and insoluble in water, although we do observe the structure deviating from the original forward mapped crystal structure. Whereas cellulose is insoluble in water it does solve in ionic liquids (IL). As Martini 3 is also capable of simulating ILs[67] we proceeded to investigate what happens, if the solvent of the above systems is changed from water to [BMIM][Cl], which cellulose is known to dissolve in.[68] Figure 7d shows the system after 2 us of simulation. In contrast to the simulation with water we do not observe a fiber like structure being formed. In addition, RDFs (Figure S5) show that the cellulose remains fully solvated. This test-case demonstrates that our Martini 3 model is in principle capable of investigating cellulose solubility.

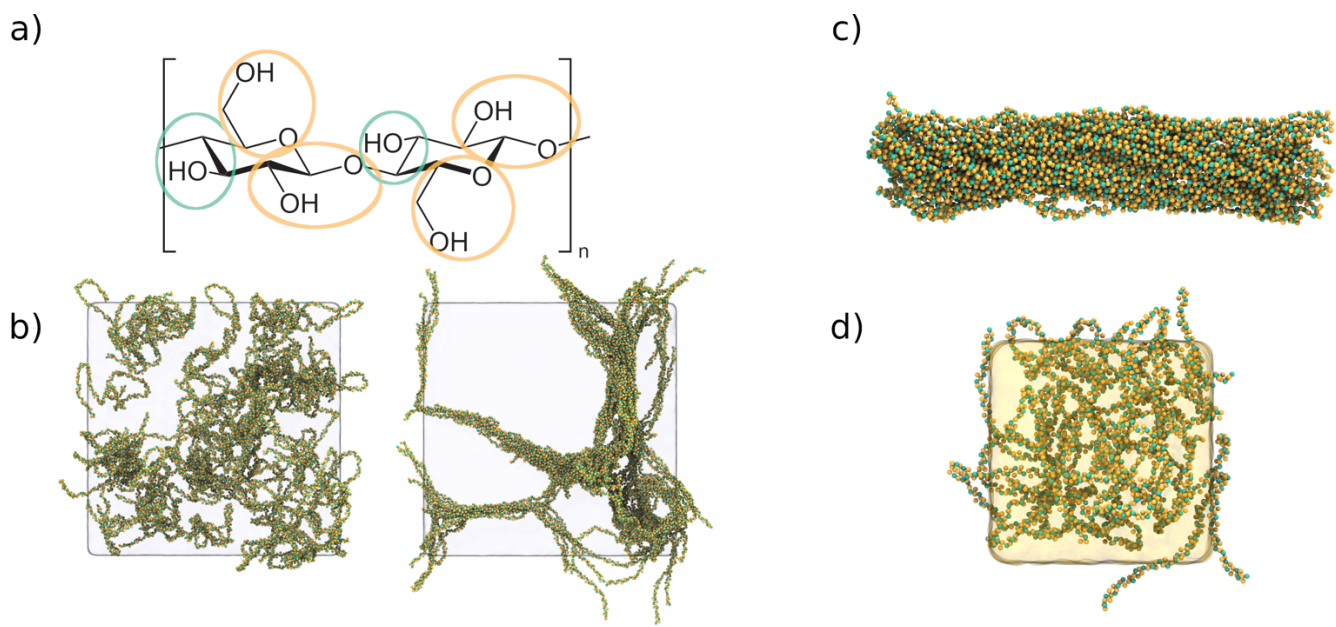

**Figure 7.** Solution properties of cellulose. a) Mapping of cellulose at the Martini level. The colors correspond to the bead-types as found in Figure 1. b) Cellulose solvated in water. Figure on the left shows 100 glucose chains with a degree of polymerization of 50, placed randomly in water. Figure on the right shows the state of the system after 2 μs of simulation. c) A cellulose Iβ fibril (36 chains with a degree of polymerization of 50) after a 2 μs simulation, solvated in water. d) Cellulose chains after 2 μs simulation, being solvated in [BMIM][Cl] ionic liquid.

### 3.4 Binding of peripheral membrane proteins to glycolipids

Lipids and lipid protein interactions in complex membranes are one of the main application areas of Martini.[69–71] However, glycolipids, which consists of a carbohydrate headgroup and various lipid tails, suffered from the same problem of excess aggregation as other sugar molecules.[18] To show that our model is transferable to lipids and proteins, we study the interaction of peripheral membrane proteins with glycolipids. In particular, we focus on bacterial Shiga and Cholera toxins, secreted by Shigella dysenteriae and Vibrio cholerae, respectively. Both toxins are associated with several human diseases e.g., diarrhea.[34] In addition, these toxins are of special interest for their applications in biophysical experiments, targeted drug delivery and cancer therapy.[72–74] Both toxins are AB$_5$ proteins that are composed of an enzymatic active A and membrane binding homopentameric B subunit. The B subunit of Cholera toxin (CTxB; Figure 8a) and Shiga toxin (STxB; Figure 8d) initiate the toxin internalization through binding to their natural receptors on the targeted cell plasma membrane: the glycolipid globotriaosylceramide (Gb3) for STxB and ganglioside (GM) for CTxB. Parameters for Gb3 and GM3 (monosialodihexosyl-ganglioside) have been designed using the presented strategy for parametrizing the carbohydrate headgroup. Lipid tail parameters were the same as in the default Martini 3 force field with adjusted linker mapping as explained in the Supporting Information, where the complete mapping for both lipids is shown (Figure S6).

First, we studied cluster formation of the GM3 lipids in a 10% POPC bilayer. Figure S7 shows the cluster-size distribution from the all-atom CHARMM36m simulation, an improved Martini 2 model proposed by Gu and coworkers[18], as well as the present model. In the AA simulation GM3 mostly exists as monomers with dimers being much less likely. Higher order clustering is almost non-existent. The same trends are captured by the fixed Martini 2 model as well as our newly proposed Martini 3 model. We conclude that both Martini models perform equivalently well, but slightly overestimate aggregation in relation to the AA simulations. However, we consider this a satisfactory result for a generic coarse-grained model. We note that the lipid tail parameters in Martini 3 are currently subject to further optimization.

Whereas realistic lipid mixtures in complex membranes were already possible to capture with the optimized Martini 2 models, binding of peripheral membrane proteins to their natural glycolipid receptor remained problematic. The Martini 2 carbohydrate model shows no specific binding sites of Gb3 to STxB. In previous studies using Martini 2, Gb3 lipids were therefore tethered to the protein via a covalent bond based on the atomistic reference structure.[75]

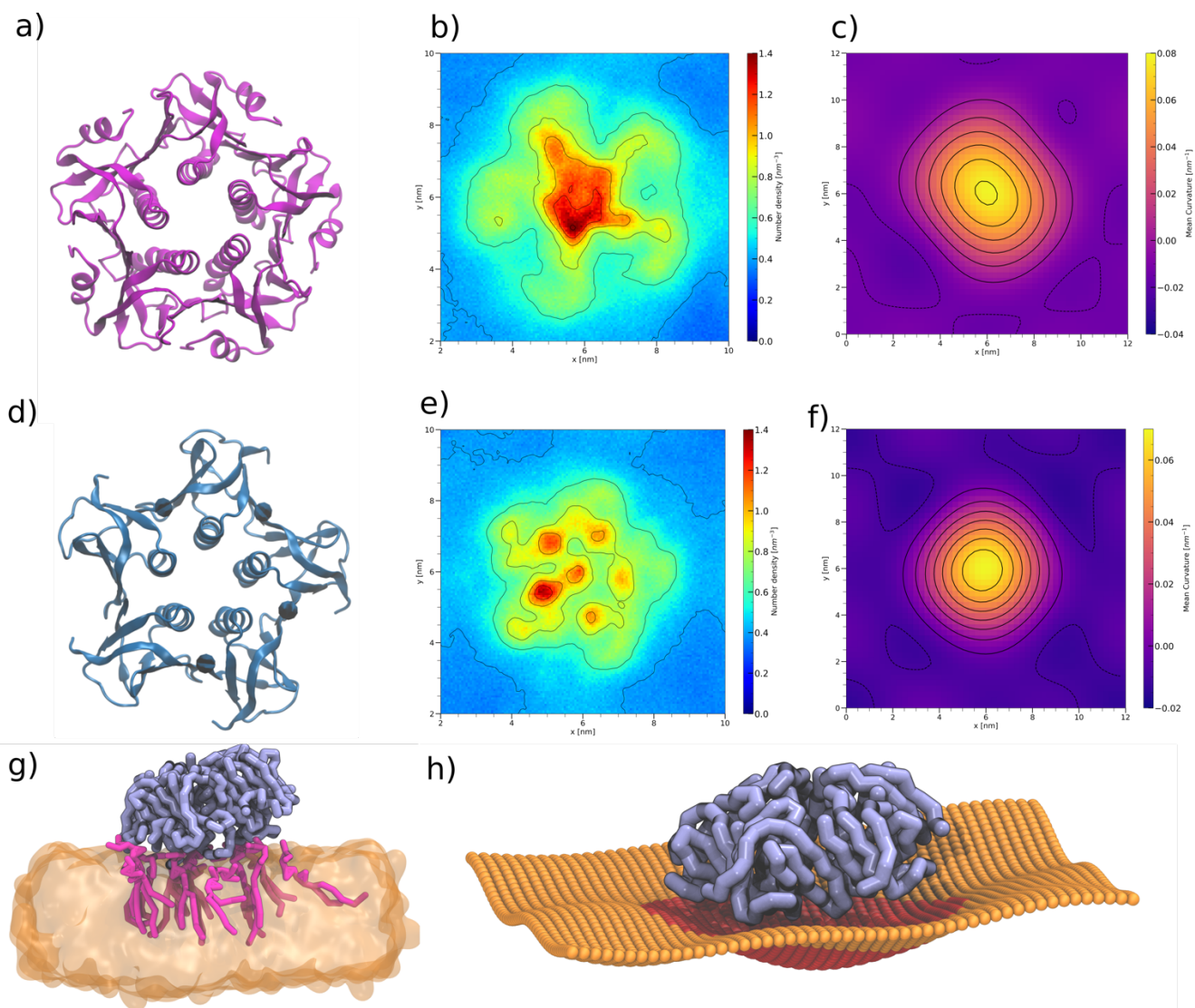

**Figure 8. Binding of peripheral membrane proteins to glycolipids. a) Rendering of Cholera toxin subunit B (CTxB) protein structure (PDB 3CHB); b) 2D lipid density map of GM3 around CTxB; c) 2D curvature plot of membrane around CTxB; d) Rendering of Shiga toxin subunit B (STxB) protein structure (PDB 2C5C); e) 2D lipid density map of Gb3 around STxB with equivalent binding sites indicated by 1-3; f) 2D curvature plot of membrane around STxB; g) CTxB (violet) bound to membrane with Gb3 lipids shown in pink and the rest of the membrane shown in transparent ;h) projected membrane surface with CTxB protein structure in the center.**

To study the process of STxB binding to Gb3 lipids with our new Martini 3 model, we simulated the system under the same conditions as the previous Martini 2 study.[75] To this end, a single STxB was placed above but not in contact with a POPC membrane, containing 10% mol fraction of Gb3. During the simulation, we observe that the protein stably binds to the membrane, not leaving during the 10 μs long simulation. A control simulation without Gb3 lipids showed no binding of STxB to the membrane over the entire course of 10μs. Similar to Shiga toxin, Cholera toxin also binds to glycolipid receptors, but in this case GM lipids including GM3. Hence, we have simulated the binding of CTxB to a POPC membrane with 10% GM3 lipids, analogously to the STxB case. Also, CTxB spontaneously binds to the membrane within a few hundreds of nanoseconds, and stays bound for the remaining 10 μs. We further analyzed binding of the glycolipids to the proteins by computing a 2D lipid density map around the centered proteins. The density maps (Figure 8b, Figure 8e) show an enrichment of glycolipids under the proteins in specific spots. For CTxB we see the most dominant binding site to be in the center of the protein and weaker binding sites in the peripheries. To our knowledge binding sites of GM3 have not been resolved for CTxB. In contrast, for Gb3 binding to STxB three distinct binding sites per monomer have been resolved experimentally by X-ray diffraction.[76] Two binding sites are found on the peripheries close to each other and a third binding site is found at the bottom of the alpha helix. In order to investigate to what extent our lipids bind in similar spots, we have computed the site-specific RDFs (see Figure S8) between the sugar parts of the glycolipids and those residues experimentally identified in binding. We see an increased probability to find a carbohydrate around each binding site indicating that binding locations appear reasonable.

Finally, we have assessed the membrane curvature induced by binding of the two proteins. Both are known to induce curvature, which is an essential step in their endocytosis.[77–79] Figure 8c and Figure 8f show the 2D mean curvature under each protein clearly demonstrating that our model can capture this behavior qualitatively. Previously the curvature for STxB bound to Gb3 has been computed from atomistic simulations ($0.034 \pm 0.004$ nm$^{-1}$)[78]. The value obtained from the CG simulations is in the same ballpark ($0.0260 \pm 0.0001$ nm$^{-1}$). The level of membrane curvature induced by binding of CTxB is further illustrated in Figure 8h.

## 4.0 – Discussion & Conclusion

In the present paper we developed a consistent strategy to parametrize arbitrary carbohydrates with the Martini 3 force field. In particular, we presented a canonical mapping scheme that decomposes arbitrarily large carbohydrates into a limited number of fragments. Bead types for these fragments have been assigned by matching atomistic volumes and free energies of transfer from water to octanol. The best bead assignment yields a mean-absolute error of about 1.3 kJ/mol compared to the experimental reference portioning data. In addition to the bead type assignment for fragments, guidelines for assigning bonds, angles, and dihedrals have been presented. These guidelines allow for a more accurate description of carbohydrate conformations than in the Martini 2 force field, and can easily be expanded to more complex carbohydrates. We showed that models obtained with this parametrization strategy are able to reproduce osmotic pressures of carbohydrate water solutions to very good accuracy. Furthermore, we demonstrated that the model differentiates correctly the solubility of the poly-glucoses dextran (water soluble) and cellulose (water insoluble). Given that the difference between both models is only a single bead-type and different bonded interactions, it speaks for the accuracy of our model being able to capture their differences. In the final test-case we illustrate that the model is applicable to glycolipids by showing that the clustering of GM3 is in good agreement with all-atom reference simulations. As last validation we analyzed the binding of peripheral membrane proteins Shiga and Cholera toxin to two glycolipid receptors. Here we found that both proteins bind to the glycolipids and induce membrane curvature as expected. Taken together, these test cases demonstrate the validity and transferability of our approach.

However, some limitations apply as well. The osmotic pressure for certain monosaccharides indicated a too high self-interaction in the high concentration regime (conc. > 1.5 molal). Therefore, simulations concerning highly concentrated solutions need to be verified carefully. Furthermore, we note that while inclusion of the TC4 virtual site greatly helps in interactions with aromatic moieties, they remain lower than observed in all-atom models. This is especially true for conformations where a stacked interaction is enforced. For example, protein binding could be influenced by this effect. Whereas part of it is an intrinsic limitation of a CG model with less degrees of freedom, further improvement can be obtained by future improvements in the protein model or even carefully revision of the Martini 3 interaction matrix. These are ongoing processes. We conclude that the rules for making carbohydrate models within Martini 3 lead to CG models that greatly improve in accuracy over Martini 2, and, at least in some aspects, are comparable to standard atomistic force fields employed in the field.

## 5.0 – Methods

### 5.1 Experimental measurements of the partition coefficients

Measurements were performed following a similar methodology as outlined in Virtanen et al[80]. Measurements were done with an UPLC-DAD-HESI-Orbitrap-MS instrument. The column in the UPLC was an Aquity BEH Phenyl (100 × 2.1 mm i.d., 1.7μm) and the mobile phase consisted of acetonitrile (A) and 0.1% aqueous formic acid (B). The elution gradient was carried out with a constant flow rate of 0.65 mL/min as follows: 0–0.1 min: 3% A; 0.1–3.0 min: 3.0–45.0% A (linear gradient); 3.0–3.1 min: 45.0–90.0% A (linear gradient); 3.1–4.0 min: 90% A; 4.0–4.1 min: 90.0–3.0% A (linear gradient); 4.1–4.2 min: 3.0% A. The ionization mode (negative/positive) of the mass spectrometer that was used for each compound depended on their ionization efficiency in either negative or positive mode; the one where each compound ionized more effectively in the test samples was then used for quantitative measurements. All measurements were done in triplicate and quantitation for each compound was done from extracted ion chromatograms (EICs) from full scan MS analysis with a specific *m/z*-range for each compound. Integrated EIC areas were converted to concentrations before partition coefficient calculations with a calibration series done with a dilution series of each compound. Both the calibration series samples and the actual $K_{ow}$ samples mass responses (integrated EIC areas) were normalized with an external standards mass response so that the possible variation in the mass spectrometers performance during the measurements and on different days could be taken into account.

## 5.2 All-atom MD simulations

All atomistic simulations were performed with GROMACS (2018.8 or 2021.5)[81].

### Simulations of mono- and dissacharides

Carbohydrates were simulated using the GLYCAM06[33] force field, the CHARMM36m[43,82] force field, and the GROMOS54a7[35,83] force field as outlined in Table S3. For each simulation a single carbohydrate was solvated in a box of water with size 2.4 x 2.4 x 2.4 nm$^3$ and simulated after short equilibration in the isobaric-isochoric ensemble at 1bar. Temperatures were fixed at 310K, 303.15K or 298.15K. Each simulation was run for at least 200 ns using the default leap-frog integrator. All bonds were restrained with the LINCS algorithm.[84] The Glycam06 and CHARMM36 simulations used TIP3P[85] as water model, whereas the GROMOS ones used the SPC model[86]. For all force fields the GROMACS specific recommended run settings were used. Itp files for the carbohydrates were obtained from the Glycam-Web and converted to GROMACS with acpype[87,88], or the CharmmGUI[43], or the automated topology builder (ATB)[35,83].

### Simulations of dextran oligomers

Single chains of dextran oligomers in water were simulated using the CHARMM36m force field[54]. Parameters and coordinates for three different degrees of polymerization (9, 14, 24) were obtained from the CHARMM-GUI[43,89]. After equilibration each simulation was run under constant temperature at 298.15K using the v-rescale[90] temperature coupling ($\tau$=1ps) with a coupling group for solvent and polymer. Pressure was kept constant at 1 bar using the Parrinello-Rahman pressure coupling algorithm ($\tau$=5ps, $\beta$=4.5x10$^{-5}$ bar$^{-1}$). The simulations for the three oligomers were run for 6$\mu$s, 3$\mu$s, and 3$\mu$s respectively. Radius of gyration and end-to-end distance were obtained from the simulation using the 'gmx polystat' tool. Distributions were subsequently computed after discarding an equilibration time.

### Simulations of lipid bilayers

All atomistic resolution lipid bilayer simulations used the CHARMM36m force field[54]. The bilayers consisted of POPC as major component and 10% glycolipids either GM3 or Gb3. Parameters and coordinates for were obtained from the CHARMM-GUI[43,89]. After equilibration each simulation was run under constant temperature at 310K using the Nose-Hoover temperature coupling ($\tau$=1ps) with a coupling group for solvent and membrane. Pressure was kept constant at 1 bar using the semiisotropic Parrinello-Rahman pressure coupling algorithm in xy an z direction ($\tau$=5ps, $\beta$=4.5x10$^{-5}$ bar$^{-1}$). The simulations were run for 2$\mu$s. Clustering of GM3 lipids was analyzed after mapping the trajectories to CG resolution with fast_forward[32]. Subsequently using the 'gmx clustsize' tool lipids were counted as being in the same cluster, if the distance between the linker beads was less than 1.4 nm.

### SASA calculations

The solvent accessible surface area was computed using the double cubic lattice method by Eisenhaber *et al.* as implemented in the GROMACS software suite (i.e. gmx sasa).[91] Instead of using the default VdW-radii for this calculation, the more recent VdW-radii proposed by Rowland and Taylor were used for the atomistic simulations.[92] For the Martini simulations the Vdw-radii were taken to be the minimum of the LJ self-interaction, which leads to three radii for the regular (0.264 nm), small (0.230 nm), tiny (0.191 nm) beads. The probe size for both atomistic and CG simulations was 0.191 nm and the SASA was averaged over at least 200 ns both for atomistic simulations and CG Martini simulations.

## 5.3 Coarse-grained MD simulations

CG simulations were performed with the Martini 3 force field[5], or Martini 2 force field[4]. Each Martini 3 simulation followed the standard simulation settings as outlined in the main publication[5] and the Martini 2 simulations followed the parameters as outlined by de Jong et al.[93], unless specified otherwise. The velocity rescaling thermostat[90] ($\tau$=1ps) and Parrinello-Rahman barostat[94] ($\tau$=12ps, $\beta$=4.5x10$^{-5}$ bar$^{-1}$) were used to maintain temperature and pressure in production simulations. GROMACS version 2021.5 was used for all simulations unless otherwise stated. Bonds within monosaccharides were constrained with the LINCS algorithm.[84]

### Free energies of transfer

All simulations pertaining free energies of transfer were carried out with the GROMACS software (version 2021.5)[81], using the stochastic dynamics integrator[95] (with inverse friction constant 1.0 ps$^{-1}$) and a time step of 20 fs. Free energies of transfer of the carbohydrates were calculated as differences between free energies of solvation in water and octanol. Solvation free energies were computed by alchemical free energy transformations as implemented in the GROMACS package. All systems for the solvation free energy in water consisted of a single carbohydrate solute molecule and 1023 Martini water beads. The systems for the octanol solvation free energies consisted of 920 octanol molecules and 80 water beads representing a saturated octanol composition.

The calculations used in total 19 non-equally spaced windows, switching only the LJ interactions as the Martini molecules considered have no partial charges. Soft-core LJ potentials were applied following the recommended values[96]. Each window was run under NpT conditions for 12 ns at 1 bar pressure maintained ($\tau$=4 ps). Temperature was maintained at 298.15K. The derivative of the potential energy was recorded every 10 steps. All the free energies of the transformation were estimated using the Bennetts-Acceptance-Ratio (BAR) method as implemented in the 'gmx bar' tool. The error reported with the calculations is the statistical error estimate. The intramolecular interactions were not switched off for both sets of simulations.

## Osmotic pressure calculations

The osmotic pressure was computed from simulations adopting the protocol originally proposed by Luo and Roux.[53] A rectangular box was created in which the solute molecules were confined in z-direction by a flat-bottomed potential to the center of the box. At a distance of 2.52078 nm from the center of the box, a harmonic force with a force constant 1000 kJ/nm$^2$ was applied to the solute molecules. The box dimensions were taken to be 10.08312 nm in z dimensions and 5.04156 nm in x and y. Previous to each run, a random configuration of solute and solvent molecules was created with polyply[64] placing solute molecules only in the center of the box and the solvent in the entire box. After energy minimization, this setup was subjected to a 10 ns equilibration using Berendsen barostat[97]. Production simulations were run for 500 ns as previously used for atomistic simulation[52] at a pressure of 1 bar. The temperature matched the temperatures reported with the experimental data-sets. Following Sauter and Grafmuller[51,55], the pressure was coupled only in z-dimensions. The osmotic pressure was computed from the trajectory by recalculating the total force exerted by the solute particles onto the flat-bottomed potential averaged over the two potentials. Subsequently that force is divided by the xy area of the box. The ensemble average as well as an error were computed from the time-series of the osmotic pressure.

## Simulation of dextran systems

Initial structures were built using polyply[64] and subsequently subjected to an energy minimization. For the scaling simulations of the oligomers first a short relaxation using the Berendsen barostat was run. Afterwards they were sampled for 3 μs using the v-rescale barostat ($\tau$=6 ps, $\beta$=4.5x10$^{-5}$ bar$^{-1}$).[98] Mixing of PEO and dextran was studied in the same fashion; however, simulations were run for 5 μs. RDFs were computed using the 'gmx rdf' tool. PEO parameters were taken from the polyply[64] library (v1.3.0). The polymer-polymer RDF was computed as an average of RDFs for each polymer separately with the other polymers as to remove the correlation induced simply by the fact that neighboring repeat units are covalently bound to each other.

## Simulation of cellulose systems

Starting configurations for the mixed state simulations in water and [BMIM][Cl] ionic liquid were generated by placing glucose chains with a DoP of 50 randomly to a large simulation volume with the 'gmx insert-molecules' tool, and then solvating with the appropriate solvent. The Martini 3 IL parameters as published earlier were used.[5,67] The starting structure of the cellulose fibril was created with the Cellulose Builder[66] after which it was solvated in the same way as the previous systems. All systems were equilibrated for 50 ns at a temperature of 310K with the system pressure controlled using the Berendsen barostat. Production simulations were run for 2 μs in the same temperature, using the Parrinello-Rahman barostat with isotropic coupling ($\tau$ = 12 ps, $\beta$ = 4.5x10$^{-5}$ bar$^{-1}$).

## Simulation of peripheral membrane protein binding

Initial structures of the lipid bilayers were built using TS2CG[75] or obtained from the atomistic simulation by mapping the bilayer and re-solvating it. All simulations were subjected to an energy minimization and equilibration. Subsequently all simulations were run under semiisotropic pressure coupling at 1bar at 310K temperature for 10μs. Protein itp files were obtained using the martinize2 code as available on GitHub. The clustering of the GM3 lipids was analyzed following the same protocol as used previously.[18] Membrane curvature was analyzed as described previously.[78] 2D density maps were computed with 'gmx densmap'. Both properties were computed as time-average over the last 7μs. Specific binding sites were analyzed using gmx rdf.

## 6.0 – Acknowledgments

We would like to thank the Center for Information Technology of the University of Groningen for their support and for providing access to the Peregrine high performance computing cluster. S.J.M. acknowledges funding from the ERC via an Advanced grant "COMP-MICR-CROW-MEM". W.P. acknowledges funding from the Novo Nordisk Foundation (grant No. NNF18SA0035142) and INTERACTIONS, Marie Skłodowska-Curie grant agreement No 847523. P.C.T.S acknowledges the support of the French National Center for Scientific Research (CNRS) and the research collaboration with PharmCADD. M.S.P.S and E.E. J. acknowledge funding by the Wellcome Trust (grant No. 203815/Z/16/Z). We would like to thank Mateusz Sikora and Philipp Schmalhorst for the early tests with the preliminary models of Martini 3 carbohydrates.

## 7.0 – Associated Content

Mono- and Disaccharide Parameters developed in this paper were deposited in a GitHub repository (https://github.com/marrink-lab/martini-forcefields) Dextran and Cellulose parameters are available from the polyply force field library starting from polyply version>1.3.2. See the following link for more details https://github.com/marrink-lab/polyply_1.0.

## 8.0 – References

# Supporting Information
# Martini 3 Coarse-Grained Force Field for Carbohydrates

Fabian Grünewald[1*], Mats H. Punt[1*], Elizabeth E. Jefferys[2], Petteri A. Vainikka[1], Valtteri Virtanen[3], Melanie König[1], Weria Pezeshkian[1,5], Maarit Karonen[3], Mark S. P. Sansom[2], Paulo C. T Souza[4], Siewert J. Marrink[1]

[1] Groningen Biomolecular Sciences and Biotechnology Institute and Zernike Institute for Advanced Materials, University of Groningen, Groningen, The Netherlands

[2] Department of Biochemistry, University of Oxford, South Parks Road, Oxford, United Kingdom, OX1 3QU

[3] Natural Chemistry Research Group, Department of Chemistry, University of Turku, FI-20014 Turku, Finland

[4] Molecular Microbiology and Structural Biochemistry, UMR 5086 CNRS and University of Lyon, Lyon, France

[5] The Niels Bohr International Academy, Niels Bohr Institute, University of Copenhagen, Copenhagen, Denmark

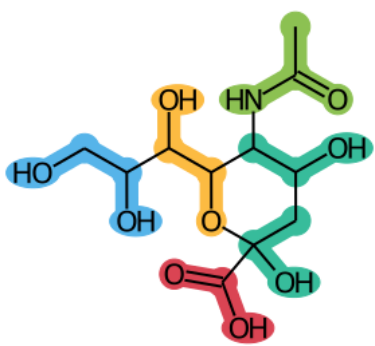

*Figure S1. Mapping of Neu5Ac carbohydrate. Colors correspond to bead types in Figure 1 and Figure 4 of the main paper.*

*Table S1. Bond lengths (nm) of $\alpha$- and $\beta$-glucose.*

| Bond | $\alpha$-D-GLC | $\beta$-D-GLC | %-diff |
|---|---|---|---|
| A - B | 0.322 | 0.330 | 2.45 |
| A - C | 0.389 | 0.409 | 5.0 |
| B - C | 0.349 | 0.344 | 1.4 |

*Table S2. Experimental partitioning coefficients.*

| Sugar | CODE | Log P | Exp | SD (n=3) | CG_final | CG_SE |
|---|---|---|---|---|---|---|
| D-glucose | GLC | -3.12 | -17.81 | 0.57 | -16.32 | 0.23 |
| D-mannose | MAN | -2.61 | -14.9 | 0.17 | -16.03 | 0.22 |
| D-galactose | GAL | -3.07 | -17.52 | 0.17 | -16.16 | 0.28 |
| N-acetylglucosamine | GlcNAc | -3.03 | -17.29 | 0.34 | -16.02 | 0.33 |
| N-acetylneuraminic acid | NMC | -4.4 | -25.11 | 0.46 | -21.39 | 0.29 |
| D-glucuronic acid | GLA | -3.26 | -18.61 | 0.11 | -18.17 | 0.31 |
| D-fucose | LFUC | -2.26 | -12.9 | 0.34 | -11.09 | 0.23 |
| L-rhamnose | LRHA | -2.26 | -12.9 | 0.23 | -11.11 | 0.24 |
| D-xylose | XYL | -2.43 | -13.87 | 0.11 | -13.06 | 0.19 |
| Inositol | INO | -3.49 | -19.92 | 0.23 | -19.42 | 0.26 |
| Trehalose* | TREH* | -3.77 | -21.52 | 0.29 | -20.61 | 0.42 |

*values taken from 10.1016/j.carres.2004.12.038

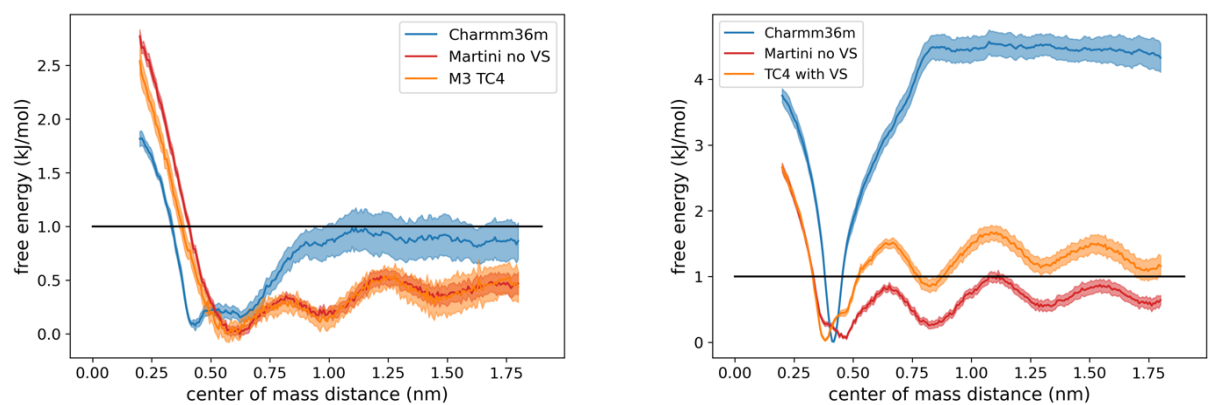

*Figure S2. Potential of mean force (PMF) of glucose - indole interaction.* We assessed the effectiveness of the virtual site (VS) by computing the potential of mean force (PMF) profiles between indole and glucose in solution without (left panel) and with orientation restraints (right panel). The orientation restraints enforce a planar interaction. We observe a reasonable agreement between Martini 3 and CHARMM36m, in case of the orientation free PMF. When planarity is enforced only the Martini model with VS shows an increased binding free energy in agreement with the atomistic simulation. However, the atomistic binding is stronger than what is recovered in Martini. Therefore, on the one hand inclusion of the VS helps in capturing aromatic interactions in general. On the other hand, in situations where a special orientation is enforced as, for example, in a tight binding pocket interactions in Martini could be underestimated.

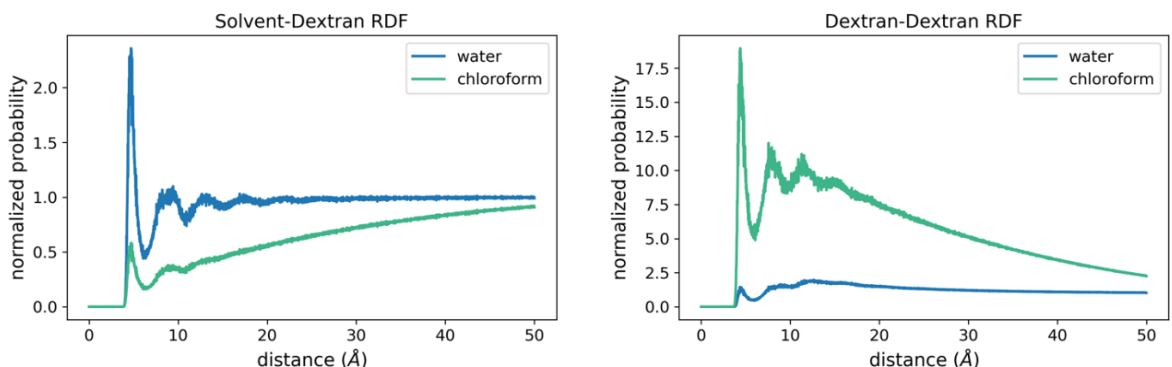

***Figure S3. Radial distribution functions of dextran with solvents and itself.*** *RDFs of the dextran beads with two different solvents, water and chloroform (left panel) were computed from 500 frames of the solution simulations. The RDF between dextran and water shows a pronounced peak at around 0.45nm corresponding to the first solvation shell. In contrast the green curve corresponding to the RDF between dextran and chloroform shows a peak which is smaller than 1 indicating that the average interactions are unfavorable as expected in a non-solvent. The self RDF between dextran, computed in the same fashion, (right panel) confirms this conclusion. Taken together these RDFs clearly demonstrate that dextran at this concentration is fully solvated in water whereas it is collapsed in chloroform.*

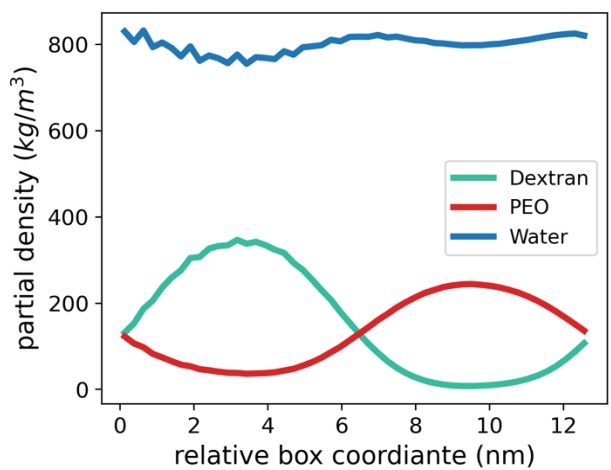

***Figure S4. Partial densities of PEO, dextran and water***. *Density profiles are shown of the ATPS of dextran and PEO in water along the z-axis of the simulation box. Clearly Dextran is depleted from the PEO phase and vice versa indicating a phase separation between the two polymers. However, the water has an almost constant partial density across the simulation box, which shows that both polymer phases are still hydrated. This corresponds to the expected density profiles for an ATPS system. We further note that water density in the dextran phase is ever so slightly less than in the PEO phase. This behavior is in qualitative agreement with experimental measurements of partitioning in higher molecular weight dextran PEO systems.*[1–3]

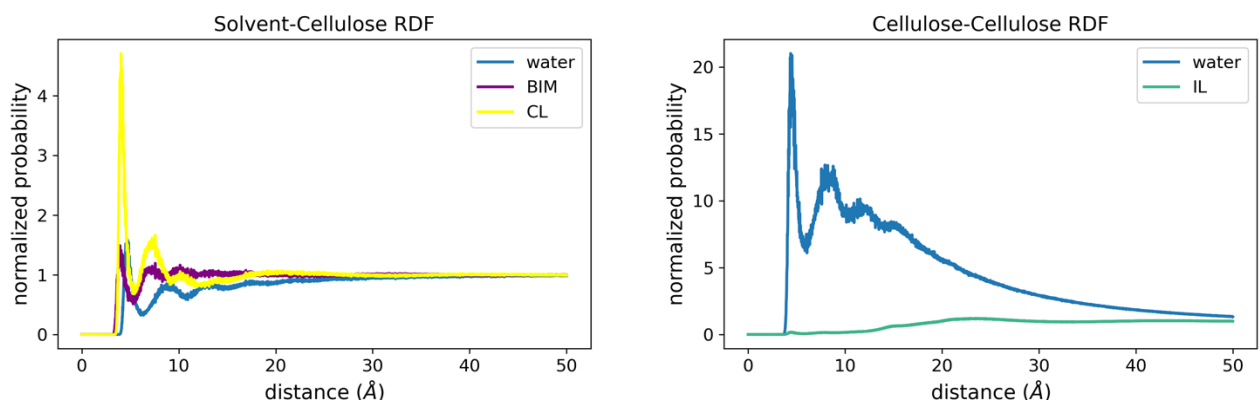

***Figure 5. Radial distribution functions of cellulose with solvents and itself.*** *RDFs of the cellulose beads with two different solvents (left panel) namely water and the ionic liquid [BMIM][CL] were computed from 500 frames of the solution simulations. The RDF between the CL of the ionic-liquid and cellulose (yellow) shows a pronounced peak at around 4.5Å corresponding to the first solvation shell. In contrast the blue curve corresponding to the RDF between cellulose and water shows a peak in that region, which is much smaller indicating that the average interactions are less favorable as expected in a non-solvent. The self RDF between cellulose, computed in the same fashion, (right panel) confirms this conclusion. Clearly the self-interaction in water (blue) is much higher than in the ionic-liquid (green). Taken together these RDFs clearly demonstrate that cellulose at this concentration is fully solvated in the ionic liquid whereas it is aggregated in water.*

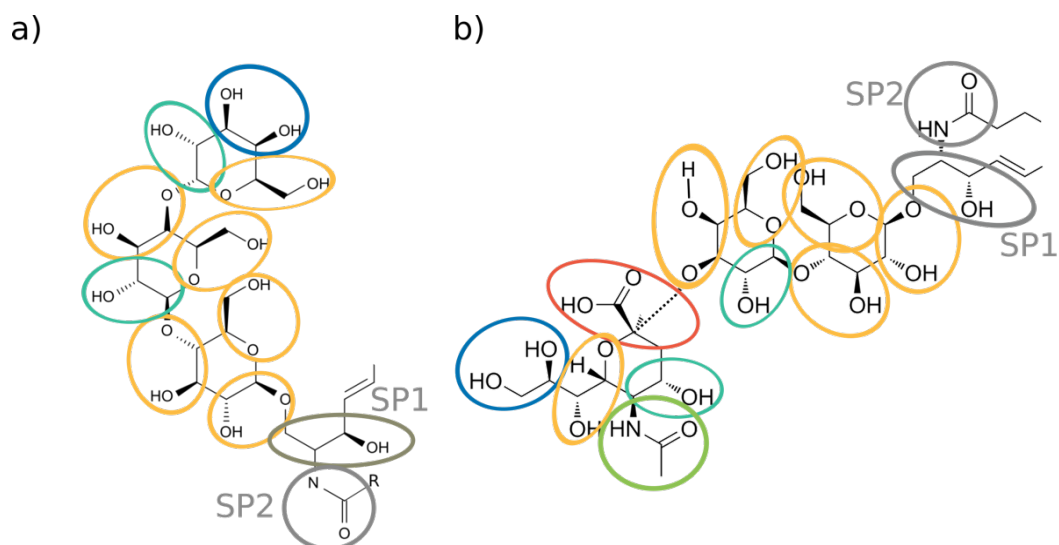

***Figure S6. Mapping of glycolipids****. Gb3 (a) and GM3 (b) with colors of the beads corresponding to Figure 1 and Figure 4 in the main manuscript. Linker beads are mapped as indicated with bead-types SP2 for the amide moiety and SP1 for the alcohol moiety. The rest of the tails is taken from the default Martini 3 paper. Note the TC4 central site has been omitted from the mapping but is part of the parameter set.*

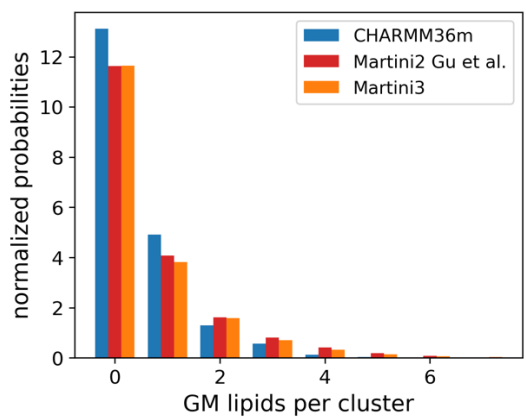

*Figure S7. Cluster size distribution for GM3 lipids in CHARMM36m, Martini 2*[4] *and Martini 3.*

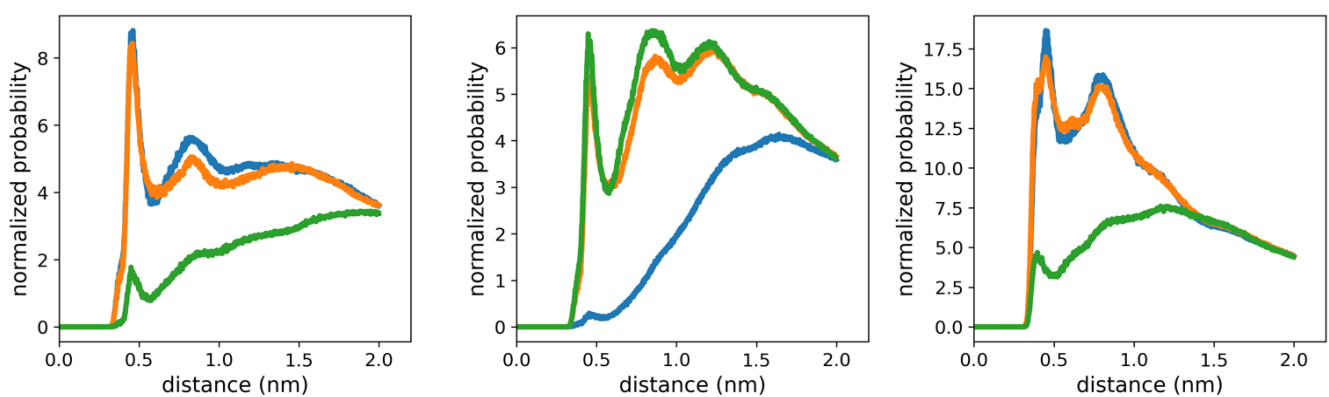

*Figure S8. Site specific RDFs of Gb3 with Shiga Toxin binding sites 1-3.* *GAL1 (orange) GAL2 (blue) and GLC3 (orange) are sequentially numbered with decreasing distance to the linker. Binding sites were taken as the whole residue at the Martini level as reported from the analysis of the X-ray crystal structure, that is, Table 3 and Table 4 in reference* [5].

*Table S3. Force-fields used for all-atom simulations of carbohydrates*

| AA SUGAR | CODE | FF |
|---|---|---|
| β-D-glucose | GLC | GLYCAM06h |
| β-D-mannose | MAN | GLYCAM06h |
| β-D-galactose | GAL | CHARMM36 |
| β-L-fucose | LFUC | CHARMM36 |
| β-L-rhamnose | LRHA | CHARMM36 |
| β-D-ribofuranose | RIBF | CHARMM36 |
| β-D-xylopyranose | XYL | CHARMM36 |
| β-D-fructofuranose | FRUF | CHARMM36 |
| Inosito | INO | GROMOS54a7 AA |
| β-D-glucuronic acid | GLA | CHARMM36 |
| β-D-N-acetylglucosamine | GYN | GLYCAM06h |
| β-D-N-acetylneuraminic acid | NMC | CHARMM36 |
| β-D-glucosamine | GCN | CHARMM36 |
| Lactose (β1,4) | LAC | CHARMM36 |
| Sucrose (α1,2) | SUCR | GROMOS54a7 AA |
| Trehalose (α1,1) | TREH | CHARMM36 |